\documentclass{jps-cp}
\usepackage{txfonts} 
\usepackage{multirow}

\title{Neutrino Experiments at J-PARC}

\author{Masahiro \textsc{Kuze}}

\inst{Department of Physics,
Tokyo Institute of Technology, 2-12-1 O-okayama, Meguro, Tokyo 152-8551, Japan}

\email{kuze@phys.titech.ac.jp}

\recdate{January 10, 2020}

\abst{Neutrino experiments at J-PARC are reviewed.  After an introduction to neutrino physics, with emphasis on neutrino oscillation, the T2K experiment at the Main Ring neutrino beam line and its peripheral experiments are introduced.  Also future prospects of the Japanese long-baseline oscillation experiments are given.  In addition, a new attempt to search for sterile neutrinos at the J-PARC MLF, JSNS$^2$ experiment, is introduced.}

\kword{neutrino, oscillation, sterile, beam, J-PARC}

\begin{document}
\maketitle

\section{Introduction on neutrino oscillation}
Neutrinos are leptons, which exist in three flavors, electron-, muon-, and tau-neutrinos.
They are subject only to the weak interaction in the Standard Model (SM) of particle physics.
They are the second ubiquitous particles next to the photon, so they played a crucial role in the early Universe.
They had been postulated as massless particles in the SM, but the neutrino oscillation phenomenon revealed that they have finite, but very small, masses.
The heaviest among them is still lighter then 1~eV, which is million times smaller than the next lightest particle, the electron.
Therefore, it is natural to think that they acquire masses with a different mechanism than Higgs coupling, by which the masses of other SM particles are accounted for.
Since neutrinos undergo only weak interactions, one needs a very intense source of neutrinos and a gigantic detector to measure them.
Consequently, the neutrino experiments need `crazy' ideas which brings fun to particle experimentalists.
Thus, the study of neutrinos is very interesting and promising,

Neutrino oscillation is a consequence of quantum mixing of neutrino states.  The weak eigenstates
($\nu_e, \nu_\mu, \nu_\tau$) are the states by which we detect neutrinos through the weak interaction, and the mass eigenstates ($\nu_1, \nu_2, \nu_3$) are the wave functions of the neutrinos traveling as a plane wave in spacetime, with definite masses ($m_1, m_2, m_3$).
These two state are not the same, and are connected by a 3$\times$3 matrix called PMNS (Pontecorvo-Maki-Nakagawa-Sakata) matrix.
It is composed as a product of three rotation matrices in three dimensions, specified by three rotation angles ($\theta_{12}, \theta_{23}, \theta_{13}$) which are called neutrino mixing angles or oscillation angles.
From the same arguments of Kobayashi and Maskawa on the quark mixing matrix, one can have a non-vanishing complex phase $e^{i\delta_{CP}}$ in the PMNS matrix, which can cause CP violation in the lepton sector, manifesting itself as a difference in neutrino and anti-neutrino oscillations.

In the approximation of two flavors (one mixing angle $\theta$), a neutrino flavor transmutes to the other flavor after traveling the distance $L$(m) with a probability $P = \sin^2 2\theta \sin^2 (1.27 \Delta m^2 L/E)$, where $E$ (MeV) is the neutrino energy and $\Delta m^2$ (eV$^2$) is the difference of the squared masses of the two mass states.
($1-P$) is called the survival probability of the initial flavor.
Therefore, if there is non-zero mixing angle and non-zero mass-squared difference (i.e. not all the neutrinos are massless!), the transmutation occurs which is called {\it neutrino oscillation}.
The amplitude of the oscillation is $\sin^2 2\theta$ and its wavelength in $L/E$ is given by $2h/(c^3\Delta m^2)$.
In the three flavor paradigm, the phenomenon is described by the three mixing angles given above, two independent mass differences ($\Delta m^2_{32} = m^2_3 - m^2_2, \Delta m^2_{21} = m^2_2 - m^2_1$, with $\Delta m^2_{31}$ being the sum of the two), and one CP-violation phase $\delta_{CP}$.

Among the three mixing amplitudes, $\sin^2 2\theta_{23}$ was measured first with atmospheric neutrinos, followed by confirmations with accelerator neutrinos.  The mixing amplitude is nearly 1.0 (close to maximal mixing), and the corresponding mass difference $|\Delta m^2_{32}|$ is around $2.5\times 10^{-3}~{\rm eV}^2$.  The next measured was $\sin^2 2\theta_{12}$, with solar neutrinos and reactor neutrinos.  It is also fairly large, about 0.8, and the corresponding mass difference $\Delta m^2_{21}$ is about  $8\times 10^{-5}~{\rm eV}^2$.  So, there is a hierarchy between the two mass squared differences; the larger one is about 30 times the smaller.  The last amplitude was measured only in 2012, by reactor neutrino and accelerator neutrino.  Surprisingly the last one is small, $\sin^2 2\theta_{13} \approx 0.1$, and the mass difference $\Delta m^2_{31}$ is almost the same as $\Delta m^2_{32}$ due to the hierarchy mentioned above.  Therefore, $\theta_{12} \approx 33^\circ, \theta_{23} \approx 45^\circ, \theta_{13} \approx 9^\circ$, and the value of $\delta_{CP}$ is still unmeasured.

According to the oscillation probability formula mentioned above, the probability becomes maximal when $1.27 \Delta m^2 L/E$ is $(2n-1)\pi/2$, where $n$ is an integer.  Since 1.27 is close to $\pi/2$, this brings a rule of thumb that you should place your detector at a distance $L$ (km) $\approx$ $E$ (GeV) / $\Delta m^2$ (eV$^2$).  Therefore, to probe the $\Delta m^2_{32(31)}$ sector with accelerator neutrino ($E \approx 1$~GeV), you get $L \approx 1/2.5\times 10^{-3}$ = 400 (km).
That is the ballpark where long-baseline accelerator experiments (T2K, MINOS, NO$\nu$A) are situated.
To probe the same sector with reactor neutrinos, with $E \approx 4$~MeV, the distance becomes 250 times smaller, i.e. 1 to 2 km, which is the distance where reactor mid-baseline experiments (Daya Bay, Double Chooz, RENO) are located.
On the other hand, to probe the $\Delta m^2_{21}$ sector, which is 30 times smaller, with the same reactor neutrinos, the distance gets 30 times larger, $L \approx 50$~km, where reactor long-baseline experiments (KamLAND, JUNO) perform measurements (actually the baseline of KamLAND is about 180 km, so the experiment is seeing the second oscillation maximum, $n$=2).

One side remark worth mentioning is that neutrino oscillation is a fascinating quantum-interference phenomenon which is observable in Earth-scale distances.  Normally quantum effects are observed only in atomic scale, but why is the scale so large in neutrino oscillation?
One then is tempted to look back the oscillation wavelength given above, $2hE/(c^3\Delta m^2)$, and decompose it to $(2h/mc)(E/mc^2)$, where $\sqrt{\Delta m^2}$ is denoted as $m$ to represent the scale of neutrino mass (which is 50~meV for $\Delta m^2_{32}$).
The first factor $h/mc$ is the Compton wavelength for a particle of mass $m$.  It is 2.4 pm for the electron, but since the neutrino mass is ten million times smaller, it is enlarged to 24~$\mu$m for a neutrino.
Furthermore, the second factor is the familiar Lorentz boost factor $\gamma = E/mc^2$.  For a 1~GeV accelerator neutrino, this boost is of the order of $10^{10}$, making overall interference wavelength of the order of 1000~km.
\begin{figure}[tbh]
\includegraphics[width=0.8\textwidth]{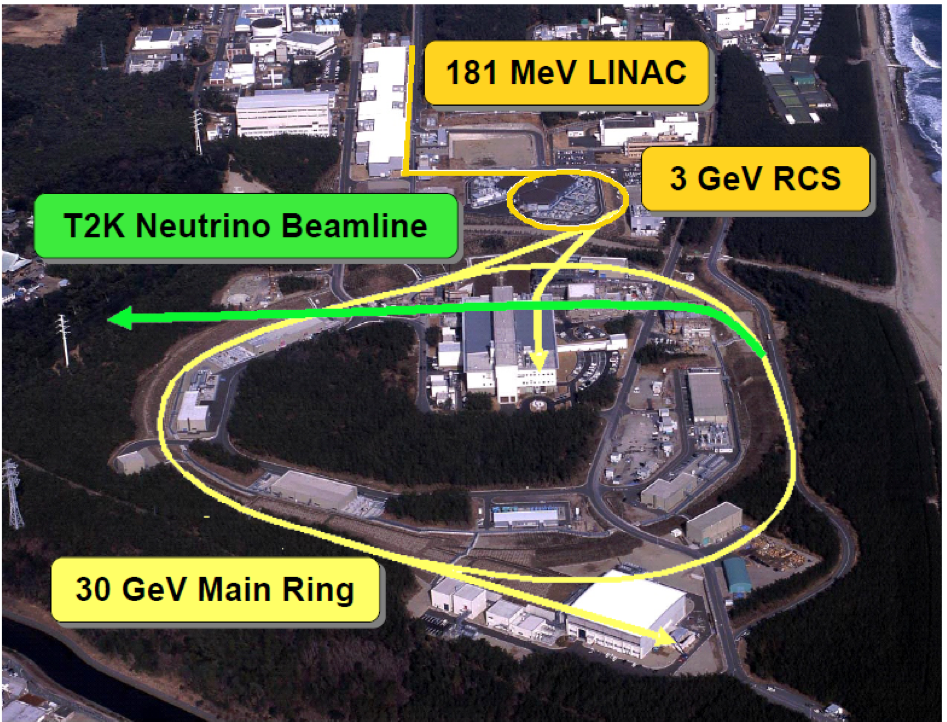}
\caption{Aerial view of J-PARC.}
\label{J-PARC}
\end{figure}

To conclude the introduction, there are various outstanding issues in neutrino physics.
All the three mixing angles have been measured (but still more precision is desired; for example, we do not know the octant of $\theta_{23}$, i.e. whether it is larger or smaller than $45^{\circ}$).
We do not know the CP violation phase $\delta_{CP}$, i.e. whether CP symmetry is violated in the lepton sector or not (while we know it is violated in the quark sector).
The method to search for CP violation consists in measuring the difference of $\nu_\mu \to \nu_e$ and $\bar{\nu}_\mu \to \bar{\nu}_e$ oscillation probabilities using accelerator neutrinos.  The detection of leptonic CP violation will shed light on the matter-antimatter asymmetry of the Universe.
Concerning the neutrino masses, both differences have been measured, and as implicitly differentiated above, the sign of $\Delta m^2_{21}$ is known to be positive from the matter effect of solar neutrinos, but only the absolute value of $\Delta m^2_{32}$ is known ($m_3$ is the one with the least mixing with $\nu_e$).  Therefore, there are two possibilities called mass ordering (mass hierarchy).  One case is $m_3 \gg m_2 > m_1$, called normal ordering (one much heavier neutrino and two lighter ones), and the other is $m_2 > m_1 \gg m_3$, called inverted ordering (two heavier and one much lighter).
The two cases will give different total mass of the neutrinos and thus affect the  evolution of matter in the early Universe.
The detection can be done using subtle matter effect of Earth in accelerator long-baseline experiments and in atmospheric neutrinos, or using subtle interference of two different-wavelength oscillations in reactor neutrino (JUNO experiment).
Note that oscillation experiments can measure only the difference of neutrino masses, but not their {\it absolute} values, i.e. how far they are from zero.
To access this information, different kinds of experiments are necessary.  One is to look for neutrino-less double-$\beta$ decays of certain isotopes, which can happen if the neutrino is a Majorana particle.
Another method is to measure the $\beta$-decay endpoint of Tritium very precisely.  KATRIN experiment is pursuing this method.
Also, cosmology can give upper limits on the sum of neutrino masses, while the neutrino oscillation results give a lower limit on the sum (currently 50~meV).

%
%
\section{T2K experiment}
\begin{figure}[tbh]
\includegraphics[width=\textwidth]{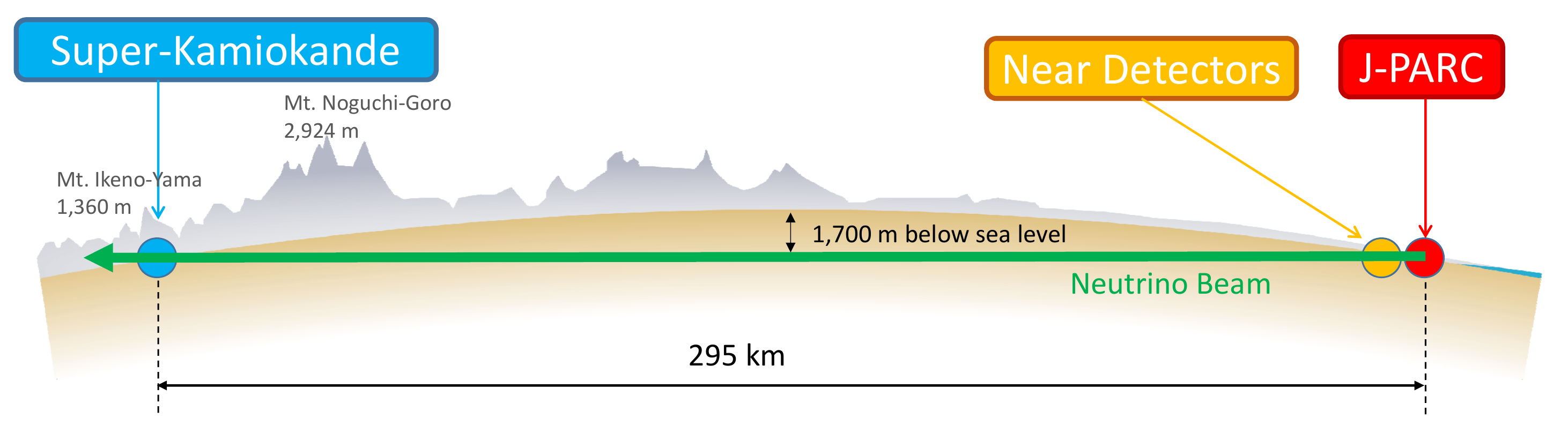}
\caption{Schematic view of the T2K experiment.}
\label{T2K-overview}
\end{figure}

\begin{figure}[tbh]
\includegraphics[width=\textwidth]{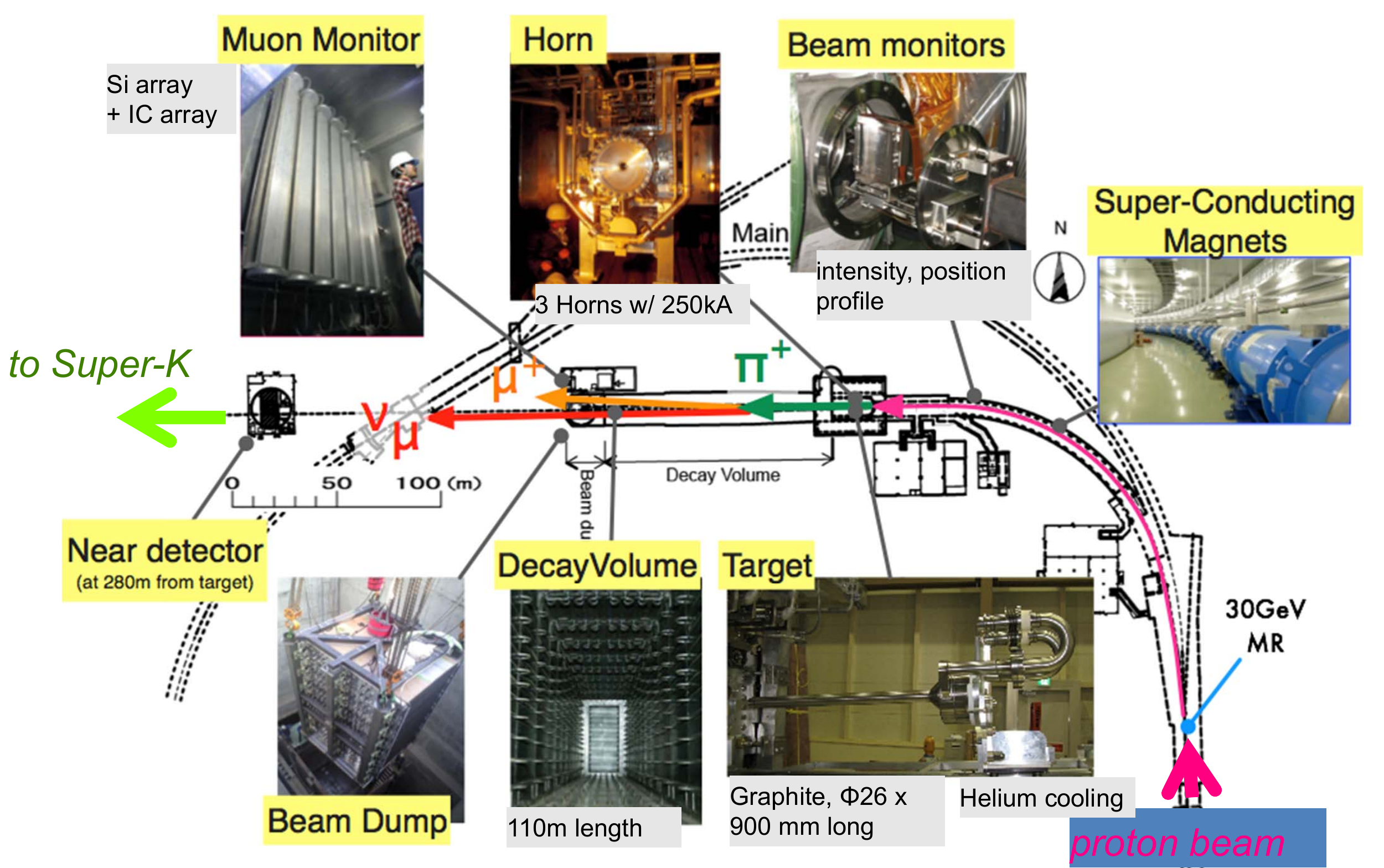}
\caption{Overview of the J-PARC neutrino beamline.}
\label{beamline}
\end{figure}

The T2K (Tokai-to Kamioka) experiment is a long-baseline accelerator-neutrino experiment in Japan.  Its baseline is 295~km from J-PARC (Ibaraki prefecture, Fig.~\ref{J-PARC}) to Super-Kamiokande (Gifu prefecture) as shown in Fig.~\ref{T2K-overview}.
At J-PARC, the neutrinos are produced by letting the fast-extraction proton beam from the 30~GeV Main Ring hit a graphite target.
Produced particles, mainly pions, are focused by three magnetic horns and brought to decay in the 100~m long decay volume filled with helium.
Therefore, the dominant component is muon neutrino from pion decay.
By switching the current polarity of the horns, pions can be charge-selected and the experiment can operate in either neutrino mode or anti-neutrino mode, with small contaminations of wrong-sign and (anti-)electron neutrinos. 
Figure~\ref{beamline} shows the components of the neutrino beam line.

\begin{figure}[tbh]
\begin{tabular}{cc}
\begin{minipage}{.5\textwidth}
\centering
\includegraphics[width=\textwidth]{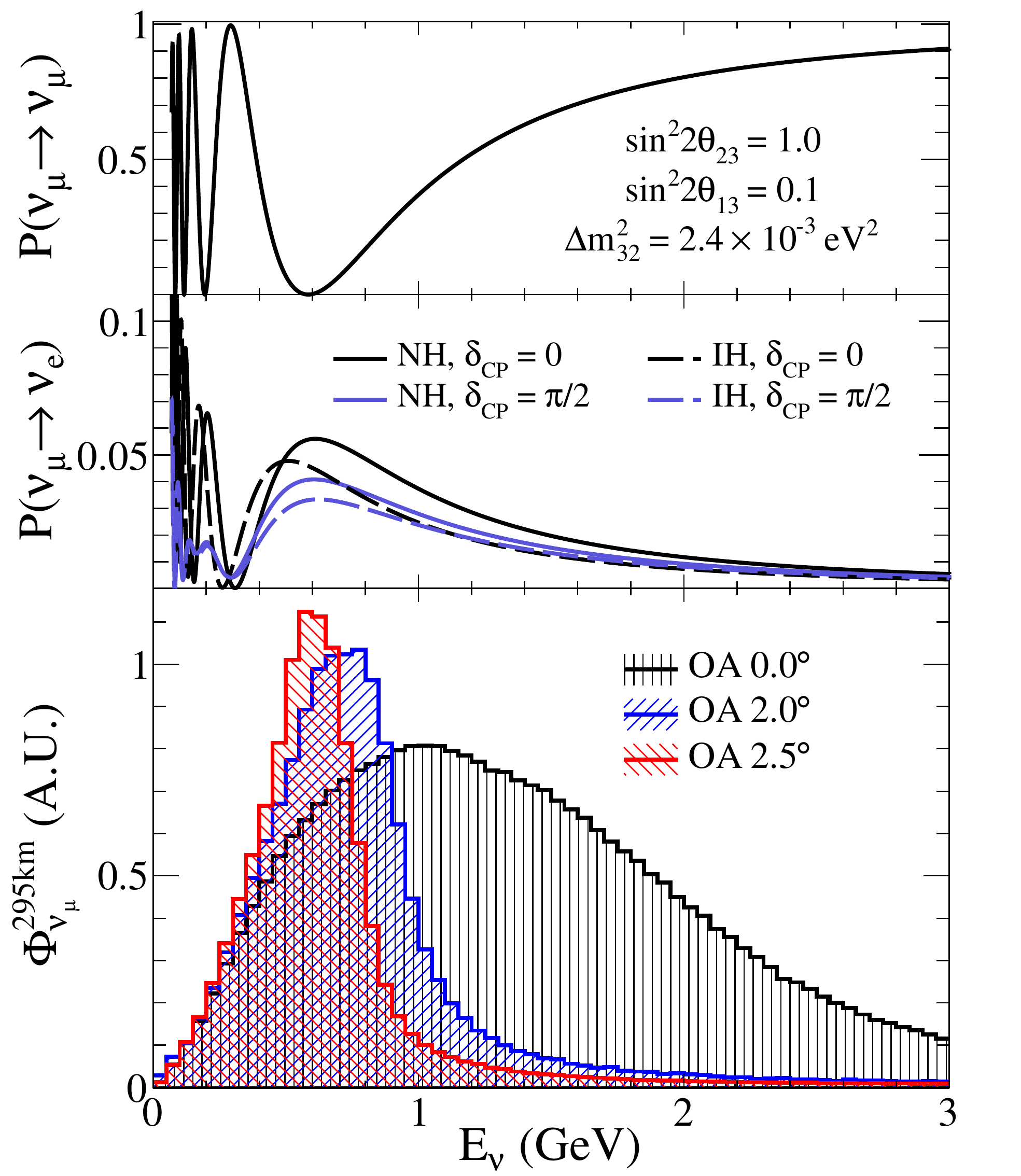}
\caption{Concept of off-axis angle neutrino beam.}
\label{off-axis}
\end{minipage}
\begin{minipage}{.5\textwidth}
\centering
\vspace{1.2cm}
\includegraphics[width=\textwidth]{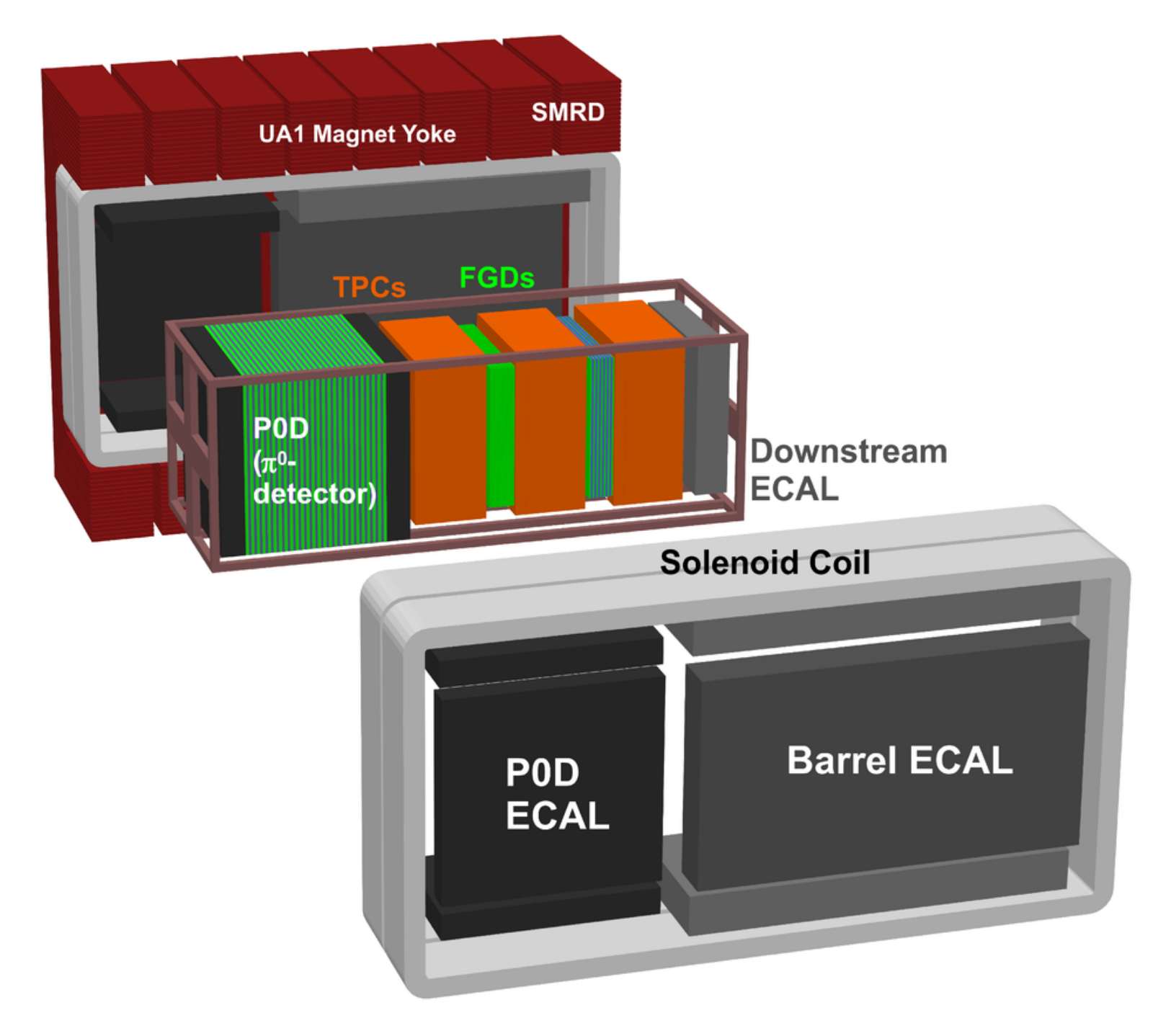}
\vspace{0.5cm}
\caption{Near detector of T2K experiment, ND280.}
\label{ND280}
\end{minipage}
\end{tabular}
\end{figure}

A unique feature of the T2K beam is that it employed for the first time the concept of off-axis neutrino beam.
Generally, the neutrino energy spectrum reflects that of the parent pions and the beam is `wide-band' along the direction of the incoming proton beam.
However, in the `off-axis' direction with increasing angles, the spectrum tends to peak at lower energies due to the two-body decay kinematics.
This way, a `narrow-band' beam is realized and in the T2K case the angle is set to 2.5 degrees giving a peak at 0.6~GeV, which matches the first oscillation maximum for $\Delta m^2_{32}$ with the 295~km baseline (Fig~\ref{off-axis}).
Not only the neutrino rate at the peak increases, but also the high-energy part of the spectrum is largely suppressed, which is a benefit for reducing background from inelastic neutral current events that could mimic electron appearance signal.
Since 2010, T2K has accumulated protons on target (POT) with increasing beam power.
At the end of 2018, it has $1.51 (1.65) \times 10^{21}$ POT of (anti-)neutrino data available for physics analysis, and the beam power reached almost 500~kW.

At 280~m from the target, T2K has a near detector, called ND280, placed at 2.5 degrees off-axis (Fig.~\ref{ND280}).
It consists of neutrino targets (plastic scintillator and water), tracking detectors and calorimeters, in a magnetic field of 0.2~T, and measures neutrino flux and spectra before the oscillation.
There is also an on-axis detector, called INGRID, to measure the neutrino beam profile (position and direction) and its stability.
The T2K far detector Super-Kamiokande, the largest tank-based water Cherenkov detector ever built, is located under 1000~m of rock (2700~m water equivalent).
It has a volume of 50 kilo-tons, of which the inner detector is 27 kilo-tons viewed by 11,129 20-inch PMTs (the outer part is viewed by different PMTs and serves as an active cosmic veto).
The particle identification (electrons and muons) is done by analyzing the shape of the Cherenkov ring
produced by the lepton exiting the neutrino interaction with water.
By synchronizing the GPS timing with J-PARC beam, the accidental background is reduced to a negligible level.
In addition to the role of T2K far detector, Super-Kamiokande has a rich physics program of atmospheric neutrino, solar, supernova and other astrophysical neutrinos, and searches for proton decay.
It underwent refurbishing work in 2018 for preparation of SK-Gadolinium(Gd) phase, and is running since January 2019 as its fifth phase (SK-V) with pure water.

\begin{table}[tbh]
\caption{Observed number of electron neutrino and anti-neutrino appearance candidates in T2K, compared with expectations for different values of $\delta_{CP},$ for $1.51 (1.65) \times 10^{21}$ POT in (anti-)neutrino mode.}
\label{t1}
\begin{tabular}{|c|c|cccc|}
\hline
 & \multirow{2}{*}{Observed} & \multicolumn{4}{c|}{Expectation} \\
 \cline{3-6}
 &  & $\delta_{CP}=-\pi/2$ &  $\delta_{CP}=0$  & $\delta_{CP}=\pi/2$  & $\delta_{CP}=\pi$ \\
\hline
$\nu_e$ candidates & 90 & 81.4 & 68.3 & 55.5 & 68.6 \\
$\bar \nu_e$ candidates & 15 & 17.1 & 19.4 & 21.7 & 19.3 \\
\hline
\end{tabular}
\end{table}
\begin{figure}[tbh]
\includegraphics[width=0.53\textwidth]{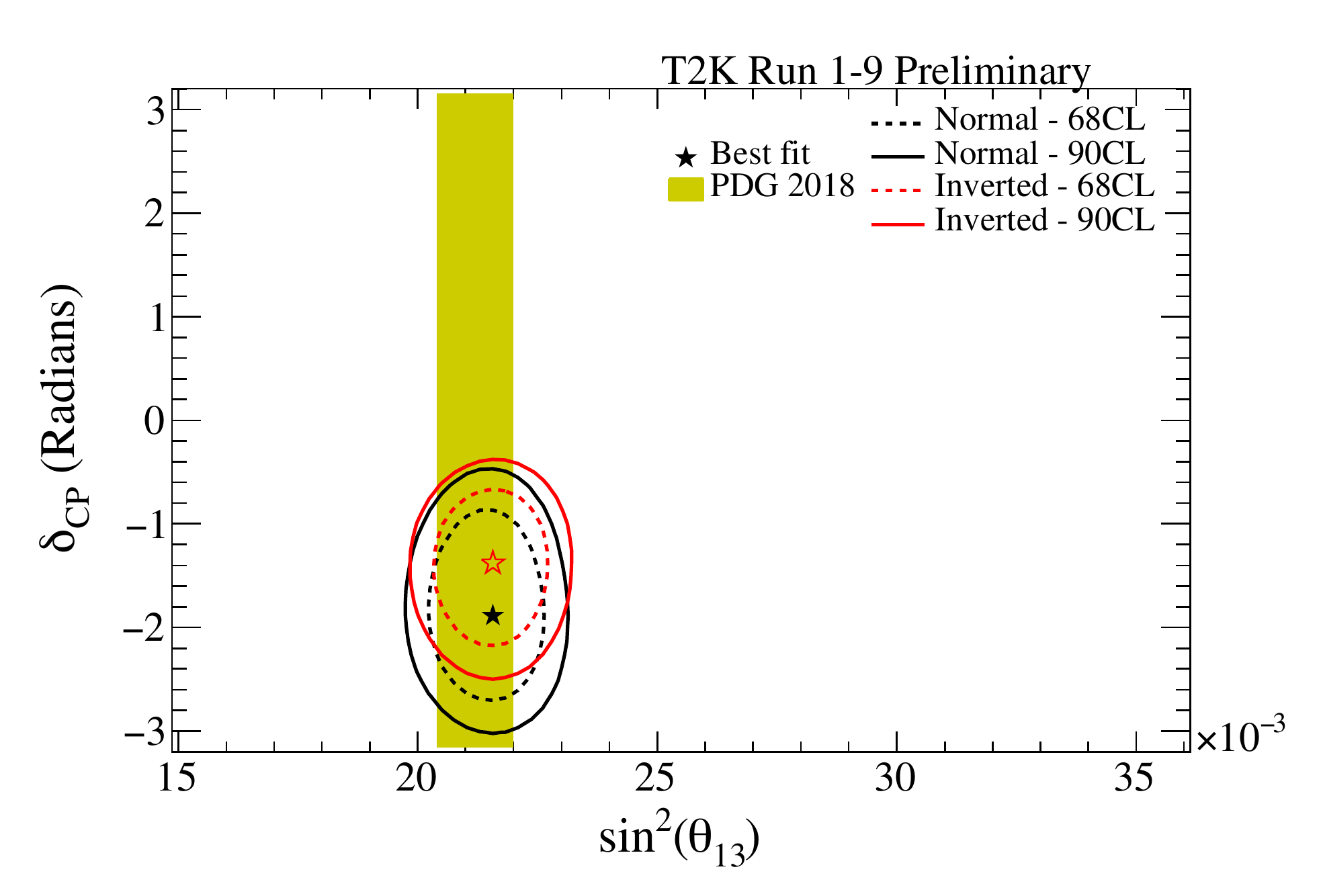}
\includegraphics[width=0.5\textwidth]{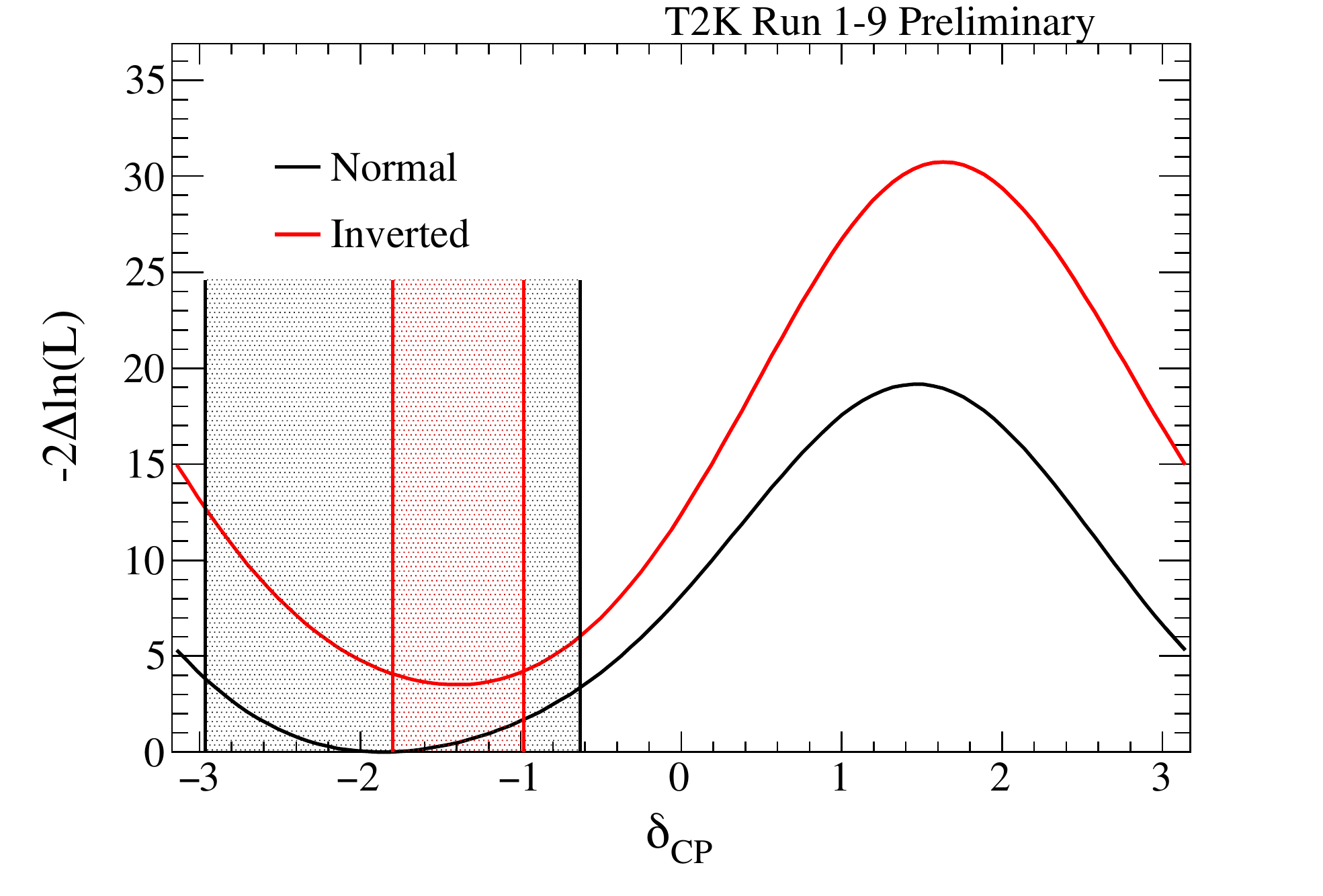}
\caption{(left) Confidence contours in the $\sin^2 \theta_{13} \-- \delta_{CP}$ plane with reactor constraint on $\theta_{13}$ (yellow band). (right) $\Delta \chi^2$ values as functions of $\delta_{CP}$.  The shaded bands show the $2\sigma$ confidence intervals.}
\label{T2KCP}
\end{figure}

Table~\ref{t1} shows the number of electron-neutrino appearance candidates in both polarity modes, compared with the expectations with different values of $\delta_{CP}$, for $1.51 (1.65) \times 10^{21}$ POT in \mbox{(anti-)}neu\-trino mode.
For $\delta_{CP}$ values close to $\pi/2$, the number of $\nu_e$ candidates should decrease and the number of $\bar \nu_e$ candidates should increase, and vice-versa in case of $\delta_{CP}$ close to $-\pi/2$.
The observation is in the direction of the latter case.
Figure~\ref{T2KCP} shows the results in terms of $\delta_{CP}$ and $\sin^2 \theta_{13}$. 
In both mass ordering cases, CP conservation ($\delta_{CP} = 0$ or $\pi$) is excluded at 2$\sigma$ level.
In addition to oscillation analysis, T2K is performing various cross section measurements that help in understanding the neutrino interactions with nuclei to minimize the corresponding systematic uncertainties in the oscillation measurement~\cite{Mahn}.

In the hall hosting ND280, new detectors have been installed recently and are taking data.
WAGASCI detector is a grid of plastic scintillator with water target, followed by muon range detector Baby MIND.
A nuclear-emulsion detector, called NINJA, accompanies them.
They are located at 1.5 degrees off-axis and their purpose is to measure the neutrino interaction cross sections and to reduce the systematic uncertainties in oscillation analysis~\cite{Yasutome}.

T2K was originally proposed for data statistics of $7.8 \times 10^{21}$ POT, and so far accumulated $3 \times 10^{21}$.
The collaboration proposes to extend the experiment (T2K-II) until ~2026, to enlarge the statistics to $15\sim20 \times 10^{21}$.
It can be realized with the upgrade of J-PARC Main Ring, which includes shorter repetition time and more protons per pulse.
The beam power will gradually increase to the goal of 1.3~MW.
With such POT, T2K has a chance to establish CP violation at 3$\sigma$ when it is maximally violated ($\delta_{CP} = -\pi/2$).
Also ND280 has a plan of upgrade for better performance~\cite{Matsubara}.

Beyond 2026, a much larger far detector, Hyper-Kamiokande is planned to be built for the T2HK experiment.
Its tank volume is 260 kilo-tons, with 190 kilo-tons of fiducial mass (compared to 22.5 kilo-tons of Super-K).
It will have 40,000 PMTs with twice the efficiency as the current one.
It is shown that Hyper-K has a discovery potential (more than 5$\sigma$) of CP violation for a wide range of $\delta_{CP}$.
The competitor is the DUNE project in US, with a similar time scale of schedule.
Hyper-K is planned to have a new intermediate water Cherenkov detector (IWCD) at around 1~km distance, in addition to the upgraded ND280.
IWCD consists of a sub-kiloton range water tank which can be moved in a vertical hole, to receive off-axis flux in a range from 1 to 4 degrees.
By combining measurements at different angles, cross sections as a function of energy can be reconstructed~\cite{Konaka}.

Furthermore, an idea exists to build a second far detector in Korea (Korean Neutrino Observatory).
The other (lower) side of the off-axis beam reaches Earth surface in the Korean Peninsula, and another off-axis measurement (e.g. at 1.5 degrees) can be made with much longer baseline (about 1100~km).
In this `T2HKK' experiment, the measurement of mass ordering will benefit from the enhanced matter effect, due to the longer baseline, enabling a determination at more than 5$\sigma$ for any value of $\delta_{CP}$.

\section{JSNS$^2$ experiment}
There is yet another neutrino source at J-PARC.  The 3~GeV Rapid Cycling Synchrotron, which is also an injector for the Main Ring, provides a high intensity proton beam to the Material and Life Science Experimental Facility (MLF), where the beam hits a mercury target acting as a spallation neutron source.
Naturally a huge number of pions are also produced in the source, which decay into muons and neutrinos.
The negative muons are mostly absorbed in the nuclei, and only positive muons stop in the material and undergo decay-at-rest (DAR), $\mu^+ \to e^+ \nu_e \bar\nu_\mu$.
Thanks to the pulsed beam structure, after waiting for 1~$\mu s$, neutrinos from meson decays are gone and a pure, isotropic source of muon DAR neutrinos is obtained.
\begin{figure}[t]
\vspace{-1.5cm}
\includegraphics[width=\textwidth]{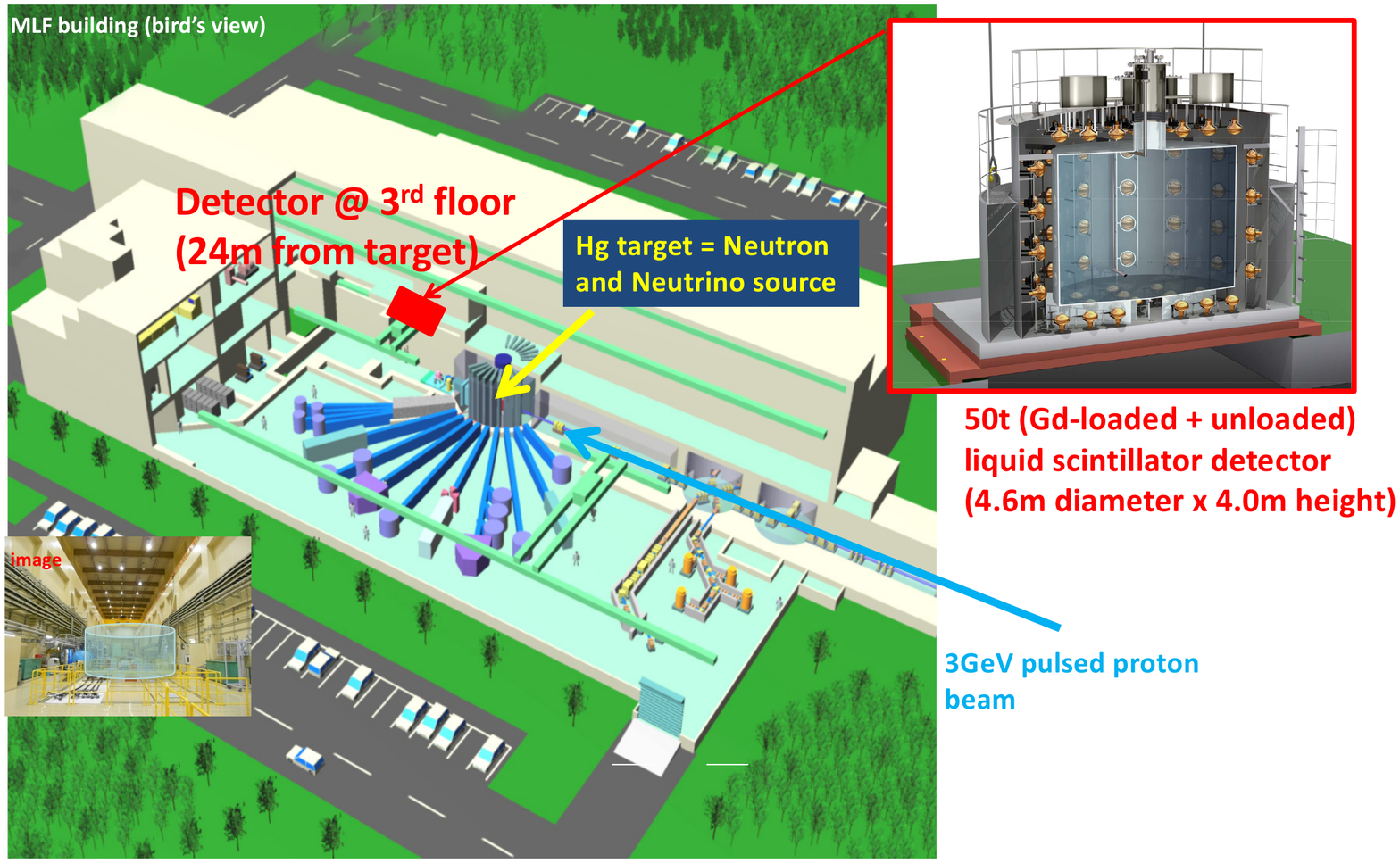}
\vspace{-1.5cm}
\caption{Schematic view of JSNS$^2$ experiment.}
\label{JSNS2exp}
\end{figure}
These DAR neutrinos are what JSNS$^2$ experiment (J-PARC Sterile Neutrino Search at J-PARC Spallation Neutron Source) will utilize~\cite{Maruyama}.
The experiment (Fig.~\ref{JSNS2exp}) searches for neutrino oscillation $ \bar\nu_\mu \to  \bar\nu_e $ at a distance of 24~m for the energy range of DAR neutrinos, i.e. 50~MeV and below.
From the `rule-of-thumb' $\Delta m^2 \approx E/L$ mentioned in the introduction, this corresponds to a search in $\Delta m^2 \approx \mathrm {1~eV}^2$ region.

Since there are only two independent $\Delta m^2$ values in the three-generation paradigm of SM, a discovery of neutrino oscillation at such large value obviously implies at least one additional degree of freedom in the mass states, i.e. a fourth-generation neutrino.
However, from the width measurement of $Z$ bosons at LEP high energy $e^+e^-$ accelerator, the species of neutrinos to which $Z$ can decay ($Z \to \nu\bar\nu$) is determined to be three, so if such a fourth neutrino exists, it is an exotic one that does not have a weak-interaction coupling.
Such exotic, hypothetical neutrinos are called {\it sterile} neutrinos.

There are several hints of experimental results that imply a neutrino oscillation at eV$^2$ range, and the most intriguing one, also long time controversial, is from the LSND experiment in US.
JSNS$^2$ is a direct test of LSND, with the same DAR neutrino and liquid scintillator detector.
It has several improvements over LSND, such as lower background rate and use of Gd-loaded scintillator.
The appearance electron anti-neutrino can be detected via $\bar\nu_e  p \to e^+ n$ interaction (Inverse Beta Decay) in which the positron annihilation creates a prompt signal and the delayed neutron capture in Gd creates a delayed signal.  This `delayed coincidence' signal greatly reduces accidental backgrounds from the environment.  The intrinsic $\bar\nu_e$ contamination from negative muon decay is suppressed below 0.5\%.
\begin{figure}[tbh]
\includegraphics[width=0.525\textwidth]{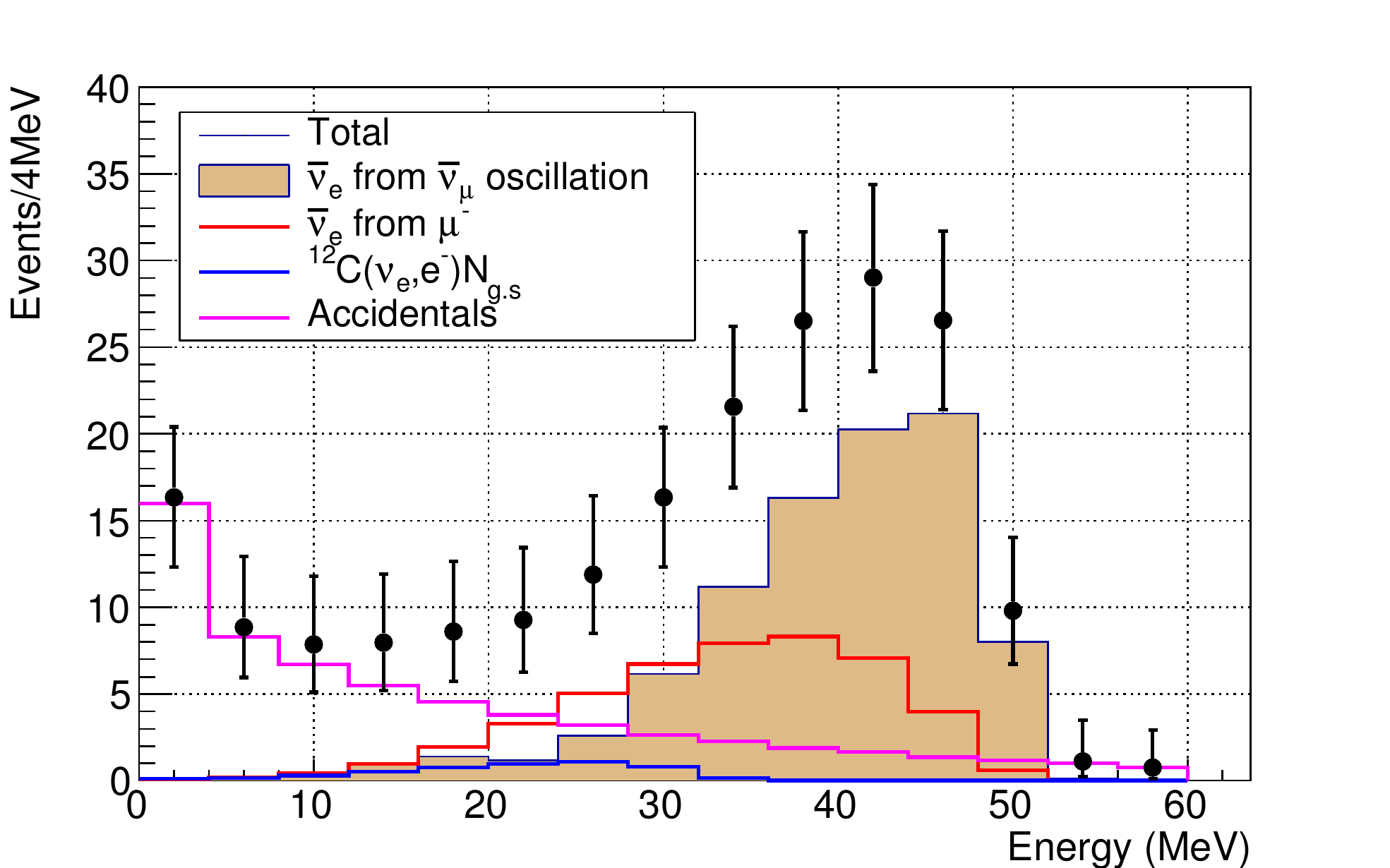}
\includegraphics[width=0.475\textwidth]{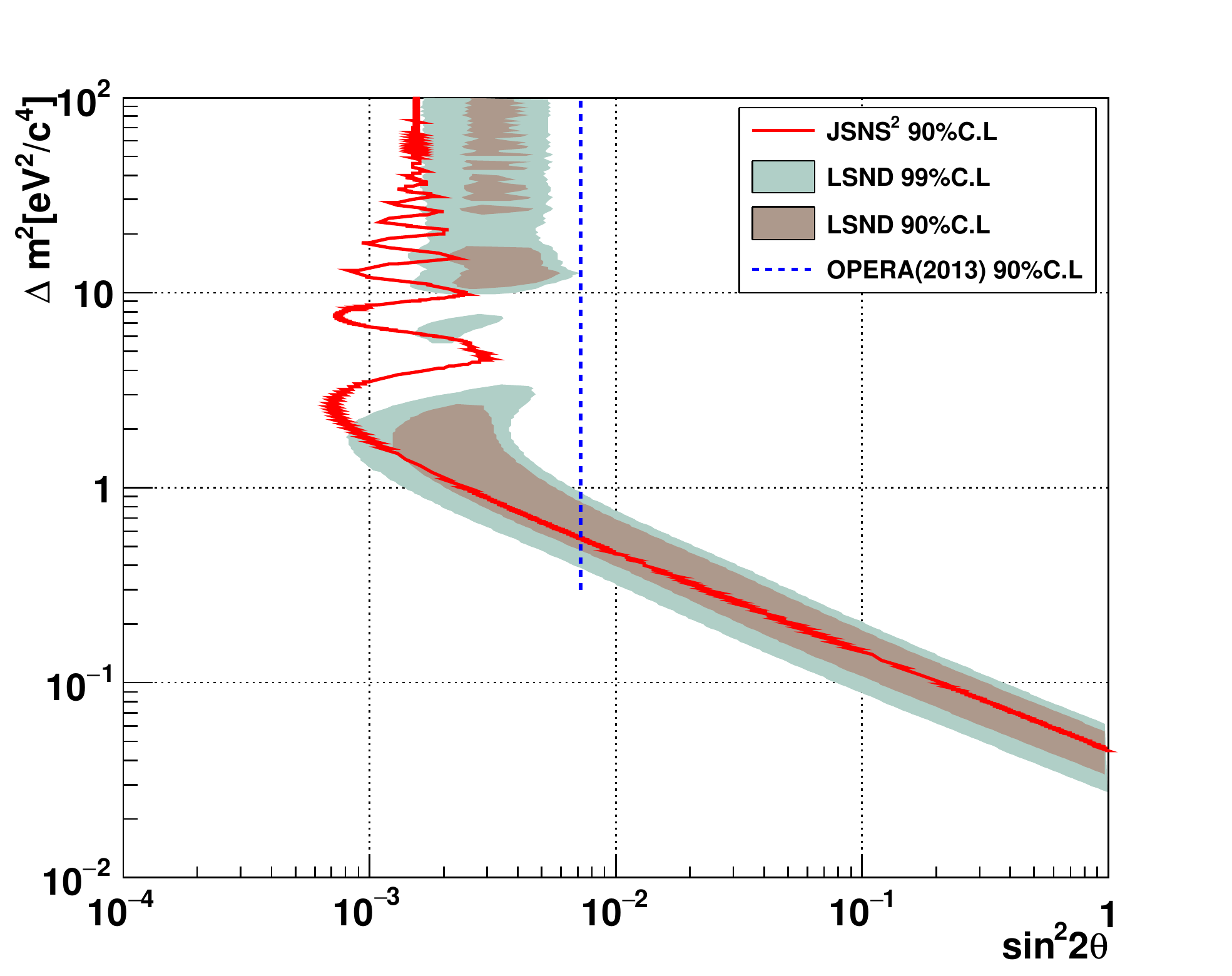}
\caption{(left) Expected signal at the JSNS$^2$ experiment for a case of $\Delta m^2=2.5~\mathrm {eV}^2$ and $\sin^2 2\theta = 0.003$ in the appearance of $\nu_e$ caused by the mixing with a sterile neutrino. Dots with error bars are sum of signal (brown shaded histogram) and backgrounds (other histograms).
(right) Sensitivity of JSNS$^2$ in the plane of $\sin^2 2\theta$ and $\Delta m^2$.
Allowed region from LSND experiment is also shown.}
\label{JSNS2}
\end{figure}

The detector is located in the upper floor of the target hall, used for maintenance work of the target station.
The neutrino target consists of 17 tons of liquid scintillator in an acrylic vessel, viewed by about 200 PMTs.
The acrylic vessel and stainless steel detector tank have been already fabricated and the collaboration is preparing the PMTs for installation.  The filling of the detector with scintillators will start soon, and the data taking is expected to start in early 2020.
Figure~\ref{JSNS2} shows the sensitivity of the JSNS$^2$ experiment.

\section{Summary}
A rich physics program of neutrino experiments at J-PARC is reviewed in this paper.
The long-baseline experiment T2K, using the neutrino beam line of Main Ring, has obtained a hint at 2$\sigma$ level that CP symmetry in the lepton sector could be violated.
This is one of the most important questions in neutrino physics in the next decade.
With a plan to increase statistics with T2K-II proposal, which includes the power upgrade of Main Ring, the experiment has a chance to establish the CP violation with 3$\sigma$.
To reduce the systematic uncertainties, T2K's near detector ND280 has a plan of upgrade, and peripheral small experiments WAGASCI and NINJA are expected to help as well.
After 2026, a much larger far detector Hyper-Kamiokande will further increase the sensitivity.
In addition, a unique experiment JSNS$^2$ is in preparation at the MLF to utilize the neutrinos from muon decay at rest, to determine the much debated existence of sterile neutrinos.

%
%
%
%
%
%


\begin{thebibliography}{9}
\bibitem{Mahn} K.~Mahn, in these proceedings.
\bibitem{Yasutome} K.~Yasutome, proceedings for NuFact 2019 conference.
\bibitem{Matsubara} T.~Matsubara, in these proceedings.
\bibitem{Konaka} A.~Konaka, in these proceedings.
\bibitem{Maruyama} T.~Maruyama, in these proceedings.
\end{thebibliography}
\end{document}